\documentclass[twocolumn]{aastex631}

\shorttitle{ICL analysis at z>0.8}
\shortauthors{Jim\'enez-Teja et al.}

\begin{document}

\title{Evidence for a redshifted excess in the intracluster light fractions of merging clusters at $z\sim 0.8$}

\email{yojite@iaa.es}

\author[0000-0002-6090-2853]{Yolanda Jim\'enez-Teja}
\affiliation{Instituto de Astrof\'isica de Andaluc\'ia--CSIC, Glorieta de la Astronom\'ia s/n, E--18008 Granada, Spain}
\affiliation{Observat\'orio Nacional, Rua General Jos\'e Cristino, 77 - Bairro Imperial de S\~ao Crist\'ov\~ao, Rio de Janeiro, 20921-400, Brazil}


\author{Renato A. Dupke}
\affiliation{Observat\'orio Nacional, Rua General Jos\'e Cristino, 77 - Bairro Imperial de S\~ao Crist\'ov\~ao, Rio de Janeiro, 20921-400, Brazil}
\affiliation{Department of Astronomy, University of Michigan, 311 West Hall, 1085 South University Ave., Ann Arbor, MI 48109-1107}
\affiliation{Eureka Scientific, 2452 Delmer St. Suite 100,Oakland, CA 94602, USA}

\author[0000-0003-2540-7424]{Paulo A. A. Lopes}
\affiliation{Observat\'orio do Valongo, Universidade Federal do Rio de Janeiro, Ladeira do Pedro Ant\^onio 43, Rio de Janeiro RJ 20080-090, Brazil}

\author[0000-0001-7399-2854]{Paola Dimauro}
\affiliation{INAF – Osservatorio Astronomico di Roma, Via di Frascati 33, 00078 Monte Porzio Catone, Italy}

\begin{abstract}
The intracluster light (ICL) fraction is a well-known indicator of the dynamical activity in intermediate-redshift clusters. Merging clusters in the redshift interval $0.18<z<0.56$ have a distinctive peak in the ICL fractions measured between $\sim 3800-4800$ \AA~. In this work, we analyze two higher-redshift, clearly merging clusters, ACT-CLJ0102-49151 and CL J0152.7–1357, at $z>0.8$, using the HST optical and infrared images obtained by the RELICS survey. We report the presence of a similar peak in the ICL fractions, although wider and redshifted to the wavelength interval $\sim 5200-7300$ \AA. The fact that this excess in the ICL fractions is found at longer wavelengths can be explained by an assorted mixture of stellar populations in the ICL, direct inheritance of an ICL that was mainly formed by major galaxy mergers with the BCG at $z>1$ and whose production is instantaneously burst by the merging event. The ubiquity of the ICL fraction merging signature across cosmic time enhances the ICL as a highly reliable and powerful probe to determine the dynamical stage of galaxy clusters, which is crucial for cluster-based cosmological inferences that require relaxation of the sample.
\end{abstract}


%

\section{Introduction}\label{sect_intro}

The intracluster light (ICL) is a diffuse light that permeates the intracluster space in galaxy clusters, composed of stars that were unbound from their progenitor galaxies by a number of mechanisms and also stars born \textit{in situ}. Some of these mechanisms are cluster-cluster mergers, major galaxy mergers with the brightest cluster galaxy (BCG), tidal stripping of luminous galaxies, shredding of dwarf galaxies, and preprocessing in infalling groups \citep[e.g., ][ and references therein]{demaio2018,jimenez-teja2023}. Although all these processes work simultaneously at all epochs, different formation pathways are more or less important at different cosmic times. According to the so called two-phase formation scenario \citep{kluge2020}, the ICL production at $z>1$ is intimately linked to the build-up of the BCG, as it is mainly fueled by stars violently unbound by mergers with this central galaxy. However, at $z<1$ the BCG and the ICL growths decouple, as other mechanisms not related with the BCG lead the ICL generation (see reviews by \citealt{contini_review2021} and \citealt{montes_review2022} for further description). \\

The nature of the ICL and its physical properties is determined by the stellar populations that compose it. \textit{In situ} star formation is found to contribute just a few percent to the total ICL budget \citep[e.g., ][]{melnick2012,gullieuszik2020}, with a few notable exceptions \citep{hlavacek-larrondo2020,barfety2022}. Generally, ICL stars were primarily hosted by progenitor galaxies in the past, so the properties of these galaxies are naturally inherited by the ICL (color, metallicity, age, etc). As a consequence, it is expected that the ICL properties evolve with redshift, as a reflection of the natural galactic evolution. Most of the observational works in the literature analyzed the ICL in nearby \citep{arnaboldi2004,lin2004,adami2005,gerhard2005,mihos2005,gonzalez2005,williams2007,coccato2011,demaio2015,edwards2016,mihos2017,jimenez-teja2019,raj2020,kluge2021,arnaboldi2022,ragusa2023}, and intermediate-redshift clusters \citep{feldmeier2002,zibetti2005,krick2006,krick2007,pierini2008,toledo2011,melnick2012,guennou2012,montes2014,adami2016,morishita2017,jimenez-teja2016,montes2018,jimenez-teja2018,zhang2019,jimenez-teja2021,furnell2021,gonzalez2021,yoo2021,deoliveira2022,dupke2022,chen2022,lee2022,montes2022,zhang2023}, where the ICL is more easily detected and analyzed. An important justification for this is the theoretical expectation of a burst of ICL at redshfit $z<0.5$ \citep{rudick2011,contini2014} which is confirmed by some observational techniques \citep{montes_review2022}. However, what happens with the ICL at higher redshift, specifically between $0.6<z<1$, is not well understood. This interval should be a transitional epoch, between the ICL growth coupled with that of the BCG and the bluer, more abundant ICL observed in clusters at lower redshifts. However, only a few observational works  attempted to study the ICL in the redshift range $0.6<z<1$ \citep[e.g., ][]{guennou2012,burke2012,burke2015,demaio2018,golden-marx2023,jimenez-teja2023}. \\

In previous works, we analyzed the ICL in a sample of low- and intermediate-redshift, massive clusters, $0.02<z<0.56$, and its relation with the dynamical state of the systems \citep{jimenez-teja2018,jimenez-teja2019,jimenez-teja2021,deoliveira2022,dupke2022}, using the most accurate method of ICL analysis to date, CICLE, \citep{jimenez-teja2016} as determined from testing against cosmological hydrodynamical simulations. We observed characteristic patterns in the ICL fraction (defined as the ratio between the ICL flux and the total light of the cluster) measured at different optical and infrared wavelengths, capable of identifying active (merging) from relaxed (passive) systems. In \cite{jimenez-teja2023}, we studied the ICL in a high redshift system, SPT-CLJ0615-574684 at $z = 0.972$, but the possible interference of a second structure in the line of sight prevented us from drawing definitive conclusions about the ICL merging signature in this earlier epoach. In this work, we perform a multiwavelength study of the ICL fraction in two higher-redshift clusters observed by the Reionization Lensing Cluster Survey \citep[RELICS, ][]{coe2019}: ACT-CLJ0102-49151 at $z=0.870$ and CL J0152.7–1357 at $z=0.833$ (ACT0102 and CL0152 hereafter, respectively). Both of them are well-know merging clusters \citep{menanteau2012,acebron2019}, so we will investigate whether the merging signature imprinted in the ICL fractions of $z<0.6$ clusters is still present at $z\sim 0.8$.\\

This paper is organized as follows. We firstly describe the main characteristics of ACT-CLJ0102-49151 and CL J0152.7–1357, along with the HST data available for them and the process of reduction in Sect. \ref{sect:data}. We describe the process followed to obtain the ICL maps in Sect. \ref{sect:ICL_maps}. Calculation of the ICL fractions and the method used for cluster membership are explained in Sect. \ref{sect:ICL_fractions}. We discuss the results of the analysis in Sect. \ref{sect:discussion} and draw the main conclusions in Sect. \ref{sect:conclusions}. Throughout this paper we will assume a standard $\Lambda$CDM cosmology with $H_0=70$ km s$^{-1}$ Mpc$^{-1}$, $\Omega_m=0.3$, and $\Omega_{\Lambda}=0.7$. All the magnitudes are referred to the AB system.

\section{Data}\label{sect:data}

The two clusters analyzed here were observed with the Advanced Camera for Surveys (ACS) in the optical and the Wide Field Camera 3 (WFC3) in the infrared (IR), both onboard the Hubble Space Telescope (HST), within the frame of the Reionization Lensing Cluster Survey \citep[RELICS,][]{coe2019}. RELICS is a multi-orbit Hubble Treasure Program that observed 41 clusters, selected by either their strong lensing power or their mass and distance.\\

ACT0102 (R.A. = 1$^{\rm h}$2$^{\rm m}$52$^{\rm s}$.50, Dec = -49$^{\circ}$14'58''0 [J2000.0], $z\sim 0.87$) is also named as  PSZ1 G297.94-67.76, PSZ2 G297.97-67.74, and ``El Gordo". This system was first detected by its Sunyaev-Zeldovich (SZ) signal \citep{marriage2011} and later confirmed through optical and X-ray data \citep{menanteau2010}. It is a clearly merging cluster, composed by a northwestern (NW) and a southeastern (SE) subclusters, with masses of $M_{200c}^{NW}=9.9_{–2.2}^{+2.1}\times 10^{14}$ and $M_{200c}^{SE}=6.5_{–1.4}^{+1.9}\times 10^{14}$ M$_{\odot}$ respectively, as determined by weak lensing \citep{kim2021}. Altogether, ACT0102 is the most massive known cluster at $z>0.8$ \citep{diego2023}. Its unrelaxed state is confirmed in X-ray by the presence of a cometary structure similar to that of the Bullet cluster \citep{molnar2015} and in radio by the detection of two radio relics and a halo \citep{lindner2014}. ACT0102 was observed in six optical and four IR bands, as detailed in Table \ref{table:data_SB}. 	\\

CL0152 (R.A. = 1$^{\rm h}$52$^{\rm m}$40$^{\rm s}$, Dec = -13$^{\circ}$57'19'' [J2000.0], $z\sim 0.833$), also known as RXJ0152.7-1357, is a massive cluster with $M_{500}=(7.8\pm 3)\times 10^{14}$ M$_{\odot}$ as estimated from X-rays \citep{sayers2013}. It was discovered in X-rays by ROSAT \citep{rosati1998,ebeling2000,romer2000} and later observed with other X-ray satellites as BeppoSAX \cite{dellaceca2000}, Chandra \citep{maughan2003,huo2004}, and XMM-Newton \citep{yuan2022}. It was also detected through its SZ signal \citep{joy2001,hilton2021}. X-rays studies show that CL0152 is composed of two main subclumps, probably bound and at the early staged of a massive merging event \citep{acebron2019}. Weak lensing analyses showed an offset of both the galaxies and dark matter centroids with respect to that of the X-rays distribution, supporting the merging scenario \citep{maughan2003,jee2005}. The X-ray peak is also found to be displaced with respect to the peak of the the SZ signal \citep{molnar2012}. Moreover, dynamical studies derived from spectroscopic measurements described the existence of several substructures, with two main ones coincident with the two X-rays subclumps \citep{demarco2005,girardi2005,jorgensen2005}. Interestingly, spectroscopy also showed that CL0152 is apparently embedded into two large-scale filaments of galaxies at $z\sim 0.837$ and $\sim ~0.844$ \citep{tanaka2006}.\\

\begin{table*}
\centering
\begin{tabular}{cccccc}
   Camera/channel & Filter & \multicolumn{2}{c}{Surface brightness limit [mag arcsec$^{-2}$]} & \multicolumn{2}{c}{ICL fractions}\\ \hline
   & & ACT0102 & CL0152 & ACT0102 & CL0152\\
\hline
ACS/WFC & F435W & $26.98\pm 0.23$ & $26.64\pm 0.23$ & ND & ND\\ 
ACS/WFC & F606W & $28.30\pm 0.06$ & ...  & $15.88\pm 1.46$ & ... \\
ACS/WFC & F625W & $27.90\pm 0.17$ & $28.06\pm 0.15$ & $8.38\pm 1.25$ & $9.77\pm 2.83$\\
ACS/WFC & F775W & $27.68\pm 0.18$ & $28.00\pm 0.09$ & $6.84\pm 0.81$ & $11.69\pm 1.77$\\
ACS/WFC & F814W & $27.65\pm 0.12$ & ... & $4.33\pm 1.78$ & ...\\
ACS/WFC & F850LP & $25.25\pm 0.04$ & $27.47\pm 0.13$ & $11.41\pm	0.93$ & $10.13\pm 1.70$\\
WFC3/IR & F105W & $27.00\pm 0.26$ & $27.18\pm 0.36$ & $21.66\pm 0.66$ & $23.44\pm 1.89$\\
WFC3/IR & F125W & $26.78\pm 0.39$ & $26.96\pm 0.49$ & $20.08\pm 1.09$ & $19.18\pm 4.79$\\
WFC3/IR & F140W & $26.84\pm 0.30$ & $26.88\pm 0.30$ & $15.92\pm 0.52$ & $16.65\pm 2.22$\\
WFC3/IR & F160W & $26.63\pm 0.24$ & $26.56\pm 0.40$ & $12.29\pm 0.54$ & $15.06\pm 3.00$\\
\hline
\end{tabular} 
\caption{ACT0102 and CL0152 data and properties. For each band, we report the $3\sigma$-surface brightness limits calculated in boxes of $10\times10$ arcsec$^2$, and the ICL fractions calculated with CICLE. We do not detect ICL in the F435W band for any of the two clusters.}\label{table:data_SB}
\end{table*}

We used the publicly available, science images reduced by the RELICS collaboration \footnote{https://archive.stsci.edu/hlsp/relics}, using the standard pipelines CALACS \footnote{{https://www.stsci.edu/hst/instrumentation/acs/software-tools/calibration-tools}} and CALWF3 \footnote{{https://www.stsci.edu/hst/instrumentation/wfc3/software-tools/pipeline}} for the optical and IR data, respectively. Final mosaics are corrected by bias, dark, flat-fielding, bias-striping, crosstalk, persistence, and charge transfer efficiency. {For the particular case of ACT0102, we found a significant variation of the background in the optical images that could compromise the ICL measurements. For completeness, we reduced the raw images of the two clusters in all bands substituting the flat-field correction derived from the standard HST pipelines by our custom one. We generated new sky-flats using the prescriptions described by \cite{borlaff2019} and reprocessed the images using them. We finally aligned and combined the individual exposures using Astrodrizzle \citep{koekemoer2002}, to generate the final mosaics with a pixel scale of 0.06 arcsec. Compared to the science images produced by the RELICS collaboration, only the optical images of ACT0102 presented significant differences, with a more homogeneous background and the low surface brightness emission better preserved. We thus used our reduced mosaics in this case.} All HST used in this paper can be found in MAST: \dataset[10.17909/hex0-qj28]{http://dx.doi.org/10.17909/hex0-qj28}.\\

{In order to test the quality of our data to detect ICL, we took the surface brightness threshold adopted by \cite{zibetti2005} to characterize the ICL, which was $\mu_r > 25$ mag arcsec$^{-2}$ in the rest-frame. To convert it to the observed surface brightness at the redshift of our clusters (we used the highest of the two redshifts for simplicity, i.e., $z\sim 0.87$), we applied the surface brightness cosmological dimming correction, the correction for the evolution the stellar population, and the k-correction to compare properly the reference and observed wavebands \citep{burke2012}. We calculated the k+e corrections using EZGAL \citep{mancone2012} and the stellar population synthesis models of \cite{bruzual2003} coupled with a \cite{chabrier2003} initial mass function, assuming an old stellar population with a formation redshift of 3 and a solar metallicity, as in previous works \citep{burke2012,burke2015,furnell2021}. Under these assumptions, the corrected surface brightness threshold at $z\sim0.87$ in the F105W band (the closest to $r$ at this redshift) is $\mu_{F105W} > 24.03$ mag arcsec$^{-2}$, which is good agreement with the surface brightness measured for clusters at $z>1$ using a completely independent method \citep{joo2023}. }Table \ref{table:data_SB} contains the $3\sigma$-surface brightness limit of the data at different bands, calculated following the procedure described in \citet{roman2020} in boxes of $10\times 10$ arcsec$^2$ (traditional angular size used for extended sources in nearby galaxies). All numbers confirm that the images are deep enough to study the ICL, with the possible exception of the F850LP image of ACT0102, which is almost $\sim 2$ mag arsec$^2$ shallower than that of CL0152. Excluding this band, the two clusters have very similar depths in all the other filters considered which, along with their close redshifts and their dynamically active state, makes them ideal targets to compare their multiwavelength ICL fractions and search for a possible merging signature on them. Both of them have low values of galactic extinction \citep[$E(B-V)=0.0086$ and 0.0126 for ACT0102 and CL0152, respectively, according to ][]{coe2019}, which guarantees a low impact on the ICL measurements. \\

\section{ICL reconstruction via galaxy fitting}\label{sect:ICL_maps}

The process followed to separate the ICL from the galactic light is based on the CHEFs Intracluster Light Estimator \citep[CICLE, ][]{jimenez-teja2018}. CICLE is a multi-galaxy fitting algorithm, that is, it fits all galaxies present in an image and later use these models to remove them and leave the ICL alone. These models are created in two dimensions, using Chebyshev rational functions and Fourier \citep[CHEFs, ][]{jimenez-teja2012} series as building blocks. CHEFs models are very flexible, capable of fitting galaxies with any kind of morphology without any a priori model constraints, and incorporate a scale parameter that permits to fit the finer substructure in the galaxies. This is specially important for high-resolution, deep imaging, as it is the case of the HST observations used in this work, since fitting using traditional, more rigid analytical profiles is prone to leave residuals that can compromise the ICL measurements after the subtraction of the models.  Especially complex is the case of the BCG since it usually has an extended halo that progressively diffuses into the ICL, making the separation between the two components not clear. CICLE identifies the transition from the BCG- to the ICL-dominated regions with a change in the curvature, that is, a difference in the slope of the projected luminous distribution of the BCG plus ICL system. This method has been tested against simulations proving a high accuracy for average configurations of the BCG+ICL system ($\sim 1\%$ of error), and a maximum error of 10\% in the most adverse cases (where the ICL was completely embedded into the BCG), in the absence of noise. Additionally, CICLE has also been compared with other techniques in a blind challenge using mock images created from four different cosmological hydrodynamical simulations {mimicking the observational characteristics of the data expected from the Vera C. Rubin Observatory after 10 years of operations \citep{brough2023}}. CICLE not only yielded the ICL fractions that were closest to the simulations, with an average difference of $-5\pm4 \%$, but also had the lowest dispersion, which proves its robustness. CICLE was also among the methods least affected by projections effects, which strengthened its reliability. \\

\begin{figure*}
\centering
\includegraphics[width=0.52\textwidth]{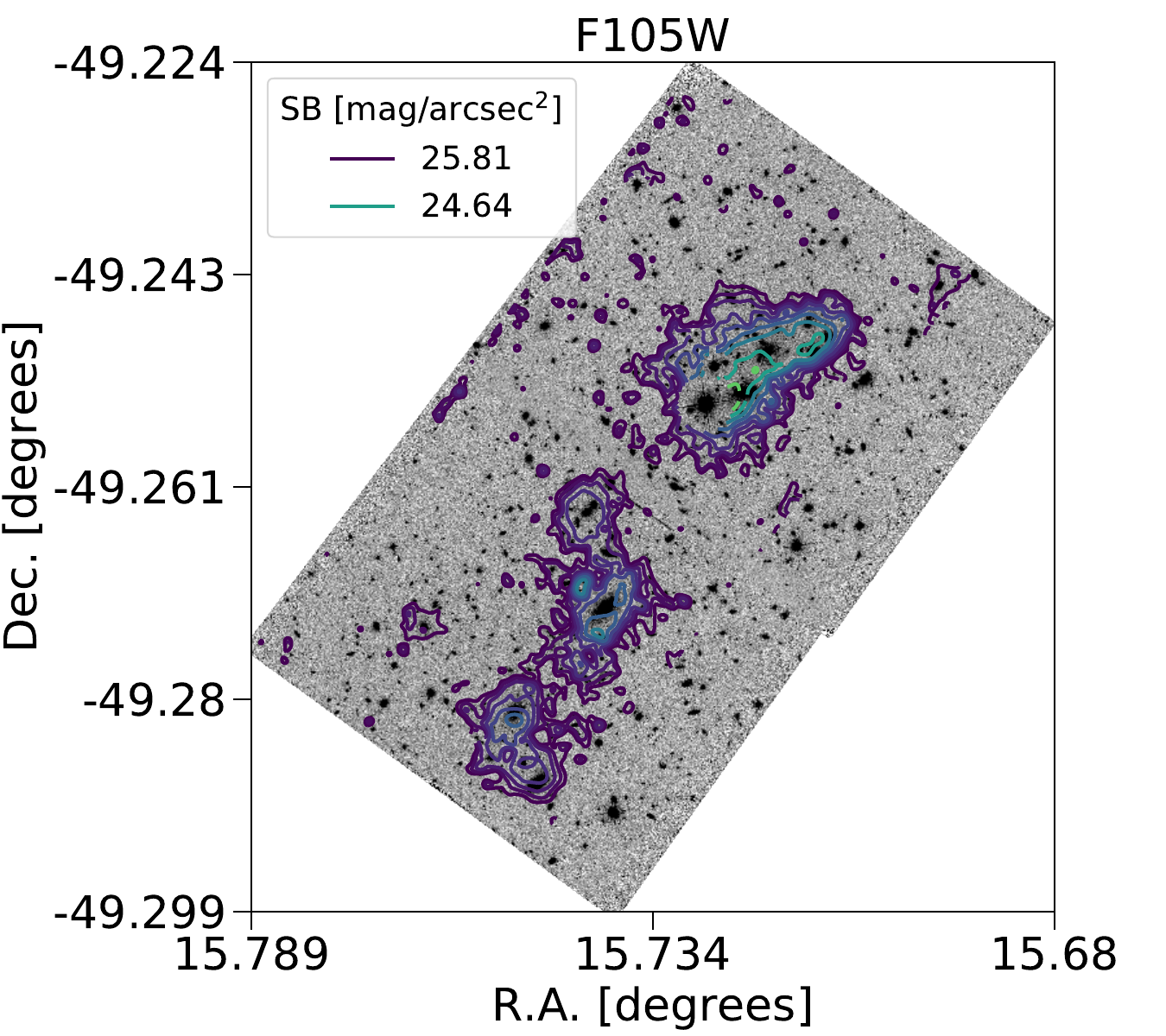}\includegraphics[width=0.52\textwidth]{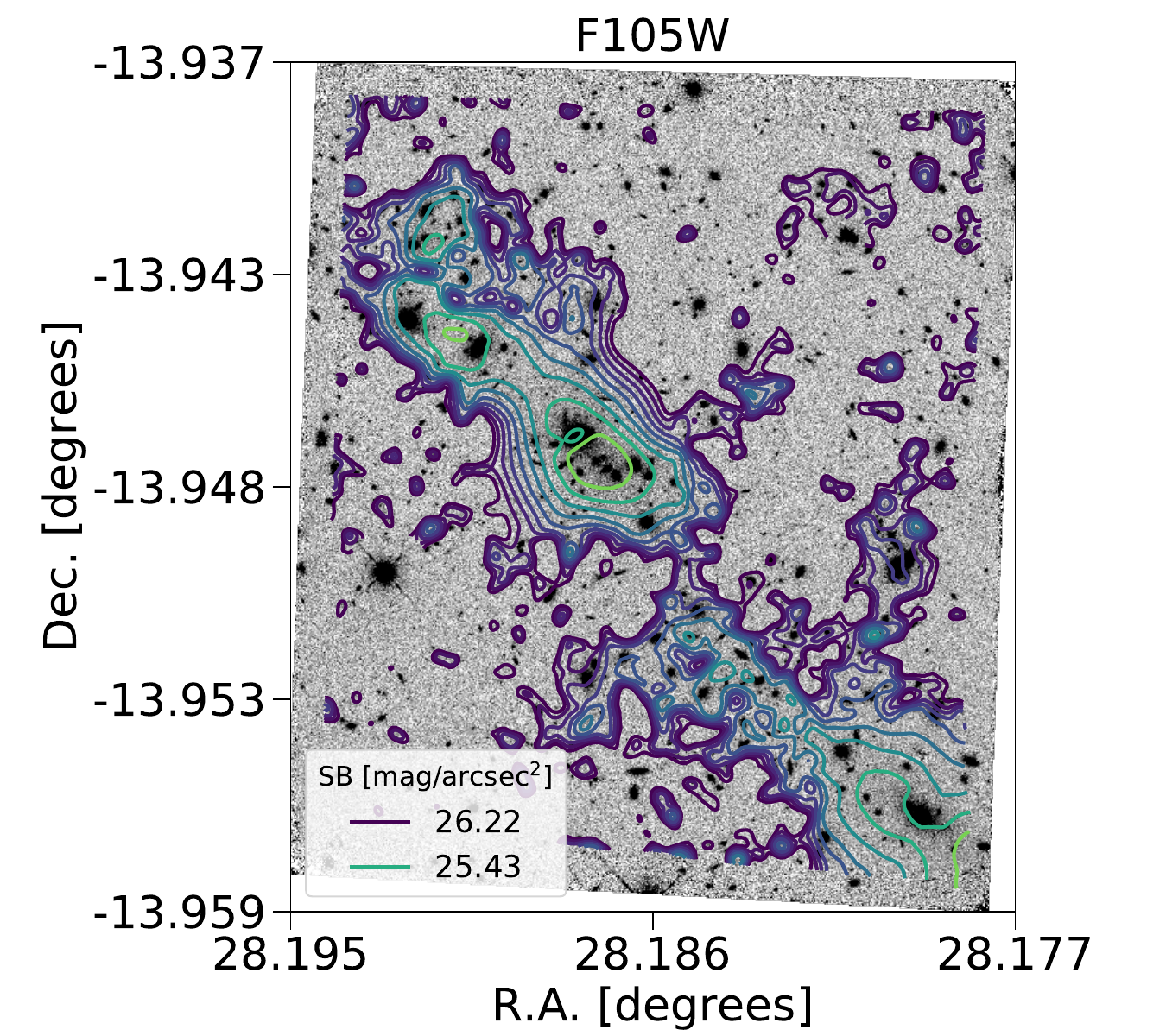}
\caption{ICL spatial distribution in the F105W band (rest-frame $\sim 5750$ \AA~at z=0.85). We plot the ICL isocontours over the original HST mosaics of ACT0102 (left) and CL0152 (right) in the F105W band.  }
\label{fig:ICLmaps}
\end{figure*}

Another crucial issue for ICL measurements is the estimation of the background. This is certainly the factor that mostly impacts the final ICL fractions, once a reasonable separation of the ICL and the galactic light is achieved. Although a first subtraction is done during the reduction of the individual exposures, final mosaics still contain a certain level of light from the sky, {despite custom sky-flats are generated specifically for each exposure (see Sect. \ref{sect:data}). NoiseChisel \citep{akhlaghi2015,akhlaghi2019} is a public software that has been proved to provide better estimations of the background than other codes \citep{borlaff2019,haigh2021,kelvin2023} and it is especially suited to detect low-surface-brightness sources and features. In this work, we used the background estimate provided by NoiseChisel during the detection routine to refine the sky subtraction.} For the sake of illustration, we show in Fig. \ref{fig:ICLmaps} the spatial distribution of the ICL of ACT0102 and CL0152 in the F105W band (rest-frame effective wavelength $\sim 5750$ \AA), superimposed over the corresponding original images of the field. In both clusters, we observe an elongated and irregular ICL that is mainly distributed in two clumps, corresponding to each one of the merging substructures identified by previous works.\\

\section{Calculation of the ICL fractions} \label{sect:ICL_fractions}

We need to identify the cluster members to re-add their CHEF models to obtain the total cluster image (galaxies plus ICL), from which we will measure the denominator of the ICL fraction. We followed a similar approach to that described in \citet{jimenez-teja2021} and \citet{jimenez-teja2023}. We applied a machine learning algorithm called Reliable Photometric Membership
(RPM, \citealt{lop20}), which assigns a probability of being a member to each galaxy based solely on its photometric information. Specifically, our training sample consisted on galaxies extracted from 18 clusters observed by the Cluster Lensing and Supernova Survey \citep[CLASH, ][]{postman2012}, spanning the redshift interval $0.0792 < z < 0.8950$. The parameters used to train the algorithm are: the local density LOG $\Sigma_5$, the difference in redshift with respect to that of the cluster $\Delta_{z\,\text{phot}}$, and the relative colors $\Delta$ (F435W-F814W), $\Delta$ (F606W-F814W), $\Delta$ (F105W-F140W), $\Delta$ (F814W-F125W), and $\Delta$ (F814W-F140W). It is important to note that these relative colors are calculated as the difference between the observed color and the mean color of the cluster galaxies. Compared to observed colors, these relative ones reduce the scatter, which is primarily induced by the large range of redshift covered by the training sample. We used spectroscopic redshifts to evaluate the quality of the final cluster membership, yielding a completeness of $93.5\% \pm 2.4\%$ and a purity of $85.7\% \pm 3.1\%$. We later applied the algorithm to a subsample of 35 clusters extracted from RELICS, among which we can find ACS0102 and CL0152. Further details on the machine-learning cluster membership can be found in \citet{jimenez-teja2021}.\\

ICL fractions are calculated as the ratio of the ICL to the total cluster light. We computed the ICL flux within the largest contour possible, outside of which the ICL submerges into the background noise. We then calculate the equivalent radius of this region to estimate the average radius up to which the ICL is detected. The ICL fraction error is composed of the contribution of the photometric error (inherent to any measurement of the flux), the geometrical error ({the error committed in the determination of the transition from the BCG- to the ICL-dominated regions plus the error due to the part of the BCG halo that extends over the ICL-dominated region}), and the flux lost due to the (in)completeness of the cluster membership. {It is important to mention that the geometrical error is calculated by testing CICLE against mock images of our two clusters, which mimic all the observed characteristics: geometry of the BCG and the ICL, pixel size, filter, and signal-to-noise ratio. Thus, the geometrical error is calculated in a similar way as in the study made by \cite{brough2023}, but adapted to the specifics of the ACT0102 and CL0152 clusters and their HST observations. }The resulting ICL fractions for both the ACT0102 and CL0152 clusters, along with their respective errors, are listed in Table \ref{table:data_SB}.\\

\section{Discussion} \label{sect:discussion}

\begin{figure*}
\centering
\includegraphics[width=\textwidth]{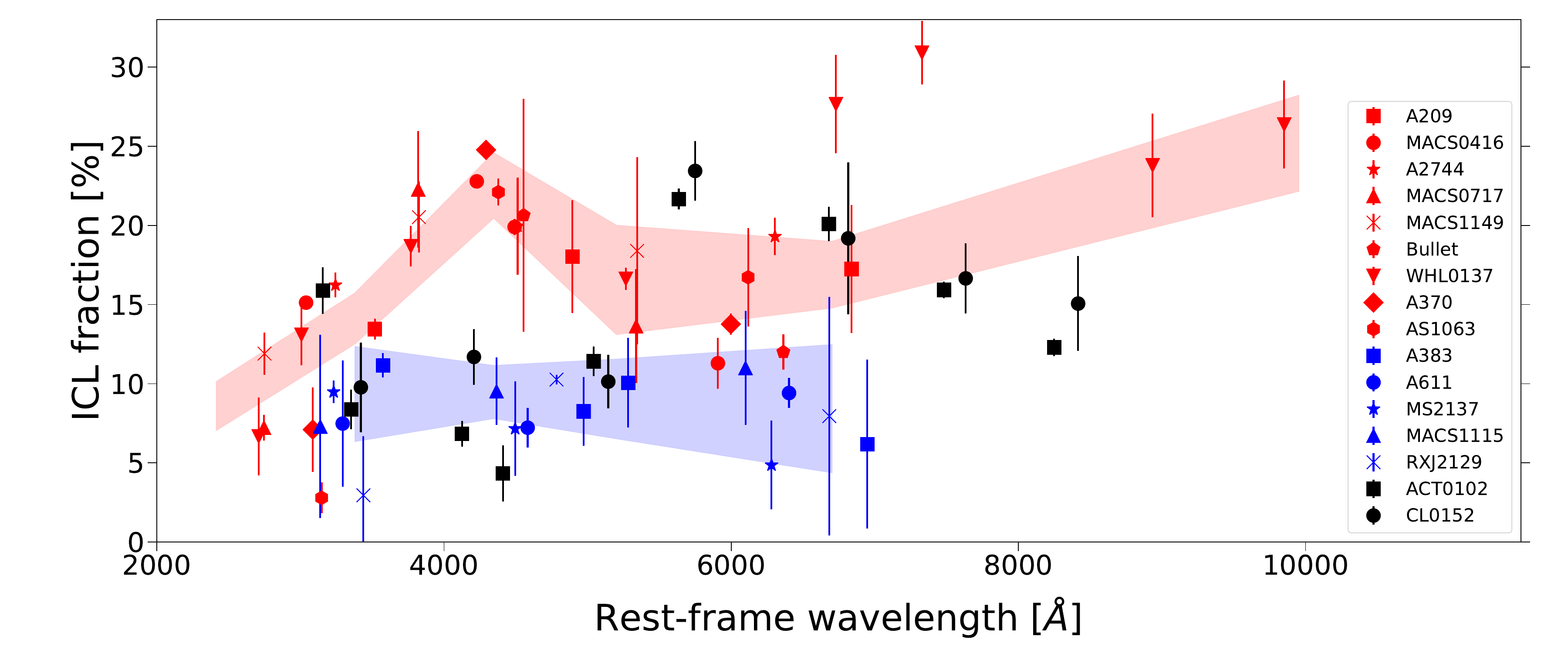}
\caption{Rest-frame ICL fractions of ACT0102 (black squares) and CL0152 (black circles), compared with those of intermediate-redshift clusters. The comparison sample spans the redshift interval $0.18<z<0.57$ and gathers the ICL fractions calculated by \citep{jimenez-teja2018,jimenez-teja2021,deoliveira2022}. Clusters represented with blue symbols are characterized by its passive dynamical stage, while those plotted with red markers are merging systems. Shadowed regions indicate the error-weighted average of the ICL fractions, for the two subsamples of relaxed and merging clusters.  }
\label{fig:ICLfractions}
\end{figure*}

In \citet{jimenez-teja2018}, we discovered that the ICL fraction carries the imprint of the recent merging activity, being significantly different for active (merging) and relaxed (virialized) clusters when  it is measured at certain specific wavelengths. This is illustrated in Fig. \ref{fig:ICLfractions}, where we show a compilation of the ICL fractions measured by \citet{jimenez-teja2018,jimenez-teja2021} and \citet{deoliveira2022}, for two subsamples of merging (red) and passive (blue) clusters. Relaxed clusters have lower and nearly constant ICL fractions in the rest-frame near-ultraviolet and visible part of the spectrum ($\sim 2000-7000$ \AA), which points to a similar stellar composition for the ICL and the most massive cluster galaxies. This is what one would expect from an ICL fed solely by passive mechanisms, such as the tidal stripping of cluster galaxies originated by dynamical friction. Contrarily, merging clusters have on average higher ICL fractions than relaxed systems, and a clear excess or peak (which we will call IE hereafter, from ICL excess) in the ICL fractions measured in the rest-frame wavelength interval $\sim 3800 - 4800$ \AA.  These extraordinarily large ICL fractions can be associated with a higher presence of younger and/or lower-metallicity stars in the ICL than in the cluster galaxies. These stars usually populate the external layers of massive galaxies \citep[e.g.,][]{denbrok2011,gonzalez-delgado2015} and would become unbound from them violently during the merging event, inducing an instantaneous rise of the ICL fraction measured at those specific wavelengths. This scenario holds for intermediate-redshift clusters ($0.18<z<0.57$), as observational evidence supports. \\

\begin{figure*}
\centering
\includegraphics[width=\textwidth]{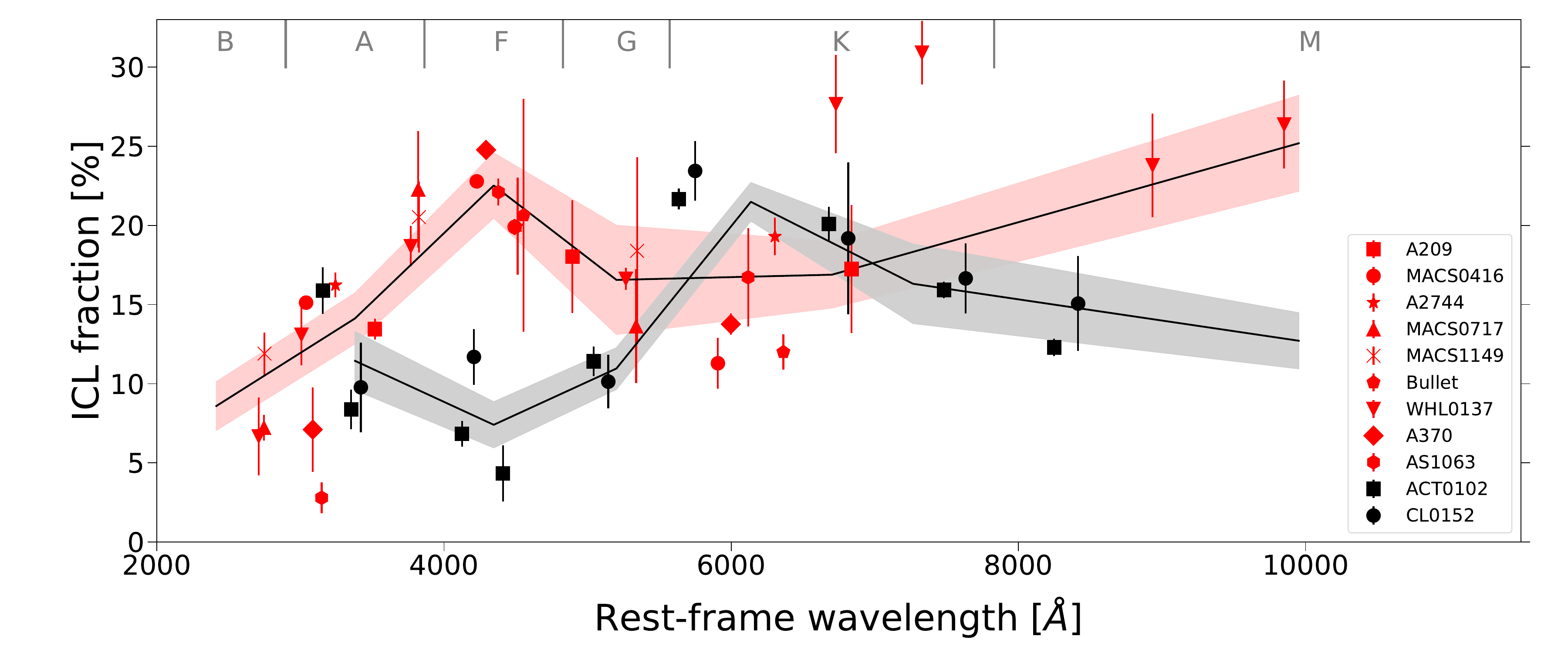}
\caption{Rest-frame ICL fractions of ACT0102 (black squares) and CL0152 (black diamonds), compared with those of dynamically active, intermediate-redshift clusters (red symbols). The electromagnetic spectrum is split into six intervals according to the emission peaks of the different stellar types, labelled with gray letters on the top. Black lines indicate the error-weighted mean for each subsample and the shaded regions represent the mean of the errors, calculated within each stellar type interval (the K-type interval has been divided in two given its width and the number of measurements that are located in it).}
\label{fig:ICLfractions_wShadows}
\end{figure*}

For this work, we selected two clusters, ACT0102 and CL015, that have similar redshifts and are well-known merging clusters \citep{acebron2019}, as confirmed by several previous studies spanning different techniques (X-rays, optical, SZ, dynamics, and weak lensing). Additionally, they have high-quality imaging data with similar observational characteristics (instrumentation, filter set, depth; see Table \ref{table:data_SB}). We plot in Fig. \ref{fig:ICLfractions} their ICL fractions with black symbols, superimposed over the measurements made for two subsamples of merging and relaxed clusters from \citet{jimenez-teja2018,jimenez-teja2021}, and \citet{deoliveira2022}. While the clusters gathered from the literature have redshifts that range between $0.18<z<0.57$, both ACT0102 and CL0152 are in a different cosmic epoch ($z\sim 0.8$) and their ICL fractions may not follow a similar trend. It is striking to observe the similarity of the ICL fractions estimated for ACT0102 and Cl0152: with the exception of the F606W (rest-frame mean wavelength $\sim 3228$ \AA) fraction of ACT0102, both clusters have low ICL fractions in the rest-frame wavelength interval $\sim 3000-5200$ \AA. The two sets of ICL fractions follow each other even more closely from $\sim 5700$ \AA~ toward redder wavelengths: they both peak in this filter and then start decreasing with a very similar slope. In summary, compared with the ICL fractions measured for intermediate-redshift clusters, we do not observe in ACT0102 and CL0152 the merging signature consisting in the characteristic IE between $\sim 3800 - 4800$ \AA) but a similar, wider peak or IE, redshifted to the wavelength interval $\sim 5200 - 7300$ \AA. For the sake of illustration and to facilitate the comparison, we show in Fig. \ref{fig:ICLfractions_wShadows} the ICL fractions found for all merging clusters (red for $0.18<z<0.57$ and gray for $z\sim 0.8$). The shadowed regions correspond to their error-weighted averages. We observe that the IE associated to the higher-redshift merging clusters appears not only redshifted, but also wider. According to the two-phase scenario of ICL formation, the production of ICL at $z>1$ is mainly driven by (major) galaxy mergers with the BCG, while at lower redshifts other mechanisms such as tidal stripping of luminous galaxies and (minor) mergers with the BCG and satellite galaxies assume the leading role. Indeed, \cite{adami2013} showed that at $z\sim 1$ there is a change in the typical mass ratio of the merging structures, so that, at this redshift approximately, the infalling groups become less massive than the accreting cluster. It is expected that this change does not occur suddenly, but gradually. Thus, at $z\sim 0.8$, the ICL still must carry the imprint of the $z>1$ major mergers mostly. As a result of them, the different stellar populations, previously hosted by the interacting galaxies, that end up unbound will mix in the ICL and produce the wider IE observed (as it comprises stars with a variety of ages and metallicities). As these major mergers primarily involves the BCG, it is expected that they produce a redder IE than that observed for intermediate-redshift merging clusters. For these intermediate-redshift merging cluster, mergers are mostly minor and, thus, they only strip stars in the outer layers of the galaxies into the ICL, which are bluer \citep{denbrok2011,gonzalez-delgado2015}, thus producing a bluer IE.\\

Observational evidence also shows that, at $z\sim 1$, there is a reversal in the star formation-density relation, so that the average star formation rate (SFR) of individual galaxies increases with the local density \citep{elbaz2007}. All galaxy mergers that happen in a dynamically active clusters, along with the fast high-SFR events must end up with a collection of post-starburst galaxies in the clusters. These post-starburst galaxies spread out star clusters of different colors that end up in the ICL, contributing to the larger width of the observed IE in the ICL fraction. A nice example of how star clusters are ejected from a post-starburst galaxy can be found in \cite{chandar2021}.\\


From the theoretical side, \citet{contini2023,contini2023b} investigated the connection between the ICL fraction and the dynamical stage using semi-analytic models of galaxy formation and found that stripping of stars would be the primary source of ICL for groups up to $log(M_{halo})\sim 10^{14}$. However, at cluster scales, they found that other mechanisms as mergers (and pre-processing in infalling groups) may share the leading role in the production of ICL, in agreement with our results.  \\

\noindent Interestingly, the ICL fractions measured at longer wavelengths ($\lambda\gtrsim 5700$ \AA) tend to decrease, contrarily to those of intermediate-redshift clusters (although Fig. \ref{fig:ICLfractions_wShadows} only shows measurements for one intermediate-redshift cluster at these redder wavelengths, WHL0137 at $z=0.56$). As ACT0102 and CL0152 are in a different cosmic time with respect to WHL0137 (with a time interval of $\sim 1.6$ Gyr), the higher presence of redder stellar populations in the ICL of this latter system with respect to the two former may obey to the natural aging of the stars and the consequent increase in the average stellar age. The main bulk of the ICL must be composed of these ancient stars, which steadily increase the ICL fractions in the rest-frame infrared part of the spectrum as cosmic time evolves. We must note that \citet{burke2012} estimated the ICL fraction of CL0152 and other clusters in the \textit{J-band} (centered at 1.2 $\mu$m) with deep VLT observations, using different isophotal levels in surface brightness to define the ICL. These authors report final fractions selecting the threshold of 22 mag arcsec$^{-2}$, for which they find an ICL fraction of $2.7\pm 0.4$\% for CL0152. If we roughly extrapolate our decreasing ICL fractions from $\lambda\gtrsim 5700$ \AA~ with a linear fit and draw the errors through jacknife resampling, we estimate a value of $1.94\pm 1.66$\% for the ICL fraction in the \textit{J}-band, consistent with the value measured by \citet{burke2012}.\\

In \citet{jimenez-teja2023}, we studied the ICL fractions of a supposedly relaxed, higher-redshift cluster, SPT-CLJ0615-574684 at $z = 0.972$. However, its rest-frame ICL fractions were not constant, but they were more consistent with those of a dynamically active cluster. The merging activity of SPT-CLJ0615-574684 was later confirmed by X-rays. However, an abnormally high ICL fraction in the near-ultraviolet was also reported, which did not follow the merging signature. As a possible group at $z\sim 0.4$ in the line-of-sight might be polluting the ICL fractions, with a special impact on the near-ultraviolet and lower-wavelength-optical regimes, we took this result with caution and we prefer to avoid a direct comparison with the ICL fractions measured for ACT0102 and CL0152. We just confirm that similar fractions are measured at $\lambda\gtrsim 5500$ \AA~ for the three clusters, the bands where the impact of the possible contamination is lower. However, at shorter wavelengths, the ICL fractions radically differ, which clear tendency to increase toward the near-ultraviolet in the case of SPT0615. As these would be the measurements that most suffered the impact of the putative group in the line of sight, we cannot draw any conclusion from the comparative of them with the two clusters analyzed in this work.\\

\section{Conclusions}\label{sect:conclusions}

We used the second and third highest-redshift clusters of the RELICS sample, ACT0102 and CL0152, to perform a multiwavelength analysis of the ICL fractions at $z\sim 0.8$. Both clusters have in common: 1) a similar cosmic epoch, 2) a clearly defined unrelaxed dynamical state, determined by several independent probes simultaneously, 3) low values of galactic extinction, and 4) HST imaging with similar observational characteristics (camera, filter set, and depth). A previous work where we analyzed the highest-redshift cluster of RELICS, SPT-CLJ0615-574684, had some uncertainties due to a possible foreground contamination by another structure in the line of sight. The main findings about ACT0102 and CL0152 are summarized in the following:
\begin{itemize}
    \item We measure strikingly similar ICL fractions for both clusters, suggestive of a common primary channel of ICL injection.
    \item ICL fractions are moderately low for the rest-frame near-ultraviolet and the blue-visible parts of the spectrum, $\sim 3000-5200$ \AA, ranging from $\sim 4$ to $16\%$.
    \item We observe a peak or excess in the ICL fractions measured in the wavelength interval $\sim 5200 - 7300$ \AA, which resembles the merging signature characteristic of intermediate-redshift clusters (which is a peak between $\sim 3800 - 4800$ \AA). This suggests that the peak in the ICL fractions, associated with dynamically disturbed clusters, is ubiquitous over cosmic time, at least up to $z\sim 0.9$. However, it does not appear always in the same part of the spectrum, but it is redshifted at earlier epochs.
    \item The peak in the ICL fractions of merging clusters at $z\sim 0.8$ is wider than that of unrelaxed systems in $0.18<z<0.56$, suggestive of a more assorted mixture of stellar populations in the higher-redshift ICL as opposed to the significant prominence of A- and F-type stars found in intermediate-redshift clusters \citep{morishita2017,jimenez-teja2018,jimenez-teja2021,deoliveira2022}.
    \item At rest-frame near-infrared wavelengths, the ICL fractions tend to decrease for high-redshift clusters, contrarily to the possible trend observed for intermediate-redshift systems. This can be explained by the steady and passive aging of the older stars, which would increase the ICL fractions measured in redder bands at lower redshift.
\end{itemize}

With this work, we reinforce the potential of the ICL fraction as indicator of the dynamical stage of massive galaxy clusters, at least up to redshift $z\sim 0.9$. Further studies would confirm or refute the ubiquity of the ICL merging signature at further redshifts and nearby systems.\\

\begin{acknowledgements}

{We thank the anonymous referee for his/her kind report and useful comments, which have largely improved the clarity and quality of this manuscript.} Y.J-T. acknowledges financial support from the European Union’s Horizon 2020 research and innovation programme under the Marie Skłodowska-Curie grant agreement No 898633 and the MSCA IF Extensions Program of the Spanish National Research Council (CSIC). Y.J-T also acknowledges the State Agency for Research of the Spanish MCIU through the Center of Excellence Severo Ochoa award to the Instituto de Astrofísica de Andalucía (SEV-2017-0709) and grant CEX2021-001131-S funded by MCIN/AEI/ 10.13039/501100011033. R.A.D. acknowledges partial support from CNPq grant 312565/2022-4. This work is based on observations taken by the RELICS Treasury Program (GO 14096) with the NASA/ESA HST, which is operated by the Association of Universities for Research in Astronomy, Inc., under NASA contract NAS5-26555. 

\end{acknowledgements}

\bibliography{YJTbibliography.bib}{}

\begin{thebibliography}{}
\expandafter\ifx\csname natexlab\endcsname\relax\def\natexlab#1{#1}\fi
\providecommand{\url}[1]{\href{#1}{#1}}
\providecommand{\dodoi}[1]{doi:~\href{http://doi.org/#1}{\nolinkurl{#1}}}
\providecommand{\doeprint}[1]{\href{http://ascl.net/#1}{\nolinkurl{http://ascl.net/#1}}}
\providecommand{\doarXiv}[1]{\href{https://arxiv.org/abs/#1}{\nolinkurl{https://arxiv.org/abs/#1}}}

\bibitem[{{Acebron} {et~al.}(2019){Acebron}, {Alon}, {Zitrin}, {Mahler}, {Coe},
  {Sharon}, {Cibirka}, {Brada{\v{c}}}, {Trenti}, {Umetsu}, {Andrade-Santos},
  {Avila}, {Bradley}, {Carrasco}, {Cerny}, {Czakon}, {Dawson}, {Frye}, {Hoag},
  {Huang}, {Johnson}, {Jones}, {Kikuchihara}, {Lam}, {Livermore}, {Lovisari},
  {Mainali}, {Oesch}, {Ogaz}, {Ouchi}, {Past}, {Paterno-Mahler}, {Peterson},
  {Ryan}, {Salmon}, {Sendra-Server}, {Stark}, {Strait}, {Toft}, \&
  {Vulcani}}]{acebron2019}
{Acebron}, A., {Alon}, M., {Zitrin}, A., {et~al.} 2019, \apj, 874, 132,
  \dodoi{10.3847/1538-4357/ab0adf}

\bibitem[{{Adami} {et~al.}(2013){Adami}, {Durret}, {Guennou}, \& {Da
  Rocha}}]{adami2013}
{Adami}, C., {Durret}, F., {Guennou}, L., \& {Da Rocha}, C. 2013, \aap, 551,
  A20, \dodoi{10.1051/0004-6361/201220282}

\bibitem[{{Adami} {et~al.}(2005){Adami}, {Slezak}, {Durret}, {Conselice},
  {Cuillandre}, {Gallagher}, {Mazure}, {Pell{\'o}}, {Picat}, \&
  {Ulmer}}]{adami2005}
{Adami}, C., {Slezak}, E., {Durret}, F., {et~al.} 2005, \aap, 429, 39,
  \dodoi{10.1051/0004-6361:20041322}

\bibitem[{{Adami} {et~al.}(2016){Adami}, {Pompei}, {Sadibekova}, {Clerc},
  {Iovino}, {McGee}, {Guennou}, {Birkinshaw}, {Horellou}, {Maurogordato},
  {Pacaud}, {Pierre}, {Poggianti}, \& {Willis}}]{adami2016}
{Adami}, C., {Pompei}, E., {Sadibekova}, T., {et~al.} 2016, \aap, 592, A7,
  \dodoi{10.1051/0004-6361/201526831}

\bibitem[{{Akhlaghi}(2019)}]{akhlaghi2019}
{Akhlaghi}, M. 2019, arXiv e-prints, arXiv:1909.11230.
\newblock \doarXiv{1909.11230}

\bibitem[{{Akhlaghi} \& {Ichikawa}(2015)}]{akhlaghi2015}
{Akhlaghi}, M., \& {Ichikawa}, T. 2015, \apjs, 220, 1,
  \dodoi{10.1088/0067-0049/220/1/1}

\bibitem[{{Arnaboldi} \& {Gerhard}(2022)}]{arnaboldi2022}
{Arnaboldi}, M., \& {Gerhard}, O. 2022, Frontiers in Astronomy and Space
  Sciences, 9, 403, \dodoi{10.3389/fspas.2022.872283}

\bibitem[{{Arnaboldi} {et~al.}(2004){Arnaboldi}, {Gerhard}, {Aguerri},
  {Freeman}, {Napolitano}, {Okamura}, \& {Yasuda}}]{arnaboldi2004}
{Arnaboldi}, M., {Gerhard}, O., {Aguerri}, J. A.~L., {et~al.} 2004, \apjl, 614,
  L33, \dodoi{10.1086/425417}

\bibitem[{{Barfety} {et~al.}(2022){Barfety}, {Valin}, {Webb}, {Yun}, {Shipley},
  {Boone}, {Hayden}, {Hlavacek-Larrondo}, {Muzzin}, {Noble}, {Perlmutter},
  {Rhea}, {Wilson}, \& {Yee}}]{barfety2022}
{Barfety}, C., {Valin}, F.-A., {Webb}, T. M.~A., {et~al.} 2022, \apj, 930, 25,
  \dodoi{10.3847/1538-4357/ac61dd}

\bibitem[{{Borlaff} {et~al.}(2019){Borlaff}, {Trujillo}, {Rom{\'a}n},
  {Beckman}, {Eliche-Moral}, {Infante-S{\'a}inz}, {Lumbreras-Calle}, {de
  Almagro}, {G{\'o}mez-Guijarro}, {Cebri{\'a}n}, {Dorta}, {Cardiel},
  {Akhlaghi}, \& {Mart{\'\i}nez-Lombilla}}]{borlaff2019}
{Borlaff}, A., {Trujillo}, I., {Rom{\'a}n}, J., {et~al.} 2019, \aap, 621, A133,
  \dodoi{10.1051/0004-6361/201834312}

\bibitem[{{Brough} {et~al.}(2023){Brough}, {Ahad}, {Bah{\'e}}, {Ellien},
  {Gonzalez}, {Jim{\'e}nez-Teja}, {Kimmig}, {Martin}, {Mart{\'\i}nez-Lombilla},
  {Montes}, {Pillepich}, {Ragusa}, {Remus}, {Collins}, {Knapen}, \&
  {Mihos}}]{brough2023}
{Brough}, S., {Ahad}, S.~L., {Bah{\'e}}, Y.~M., {et~al.} 2023, \mnras,
  \dodoi{10.1093/mnras/stad3810}

\bibitem[{{Bruzual} \& {Charlot}(2003)}]{bruzual2003}
{Bruzual}, G., \& {Charlot}, S. 2003, \mnras, 344, 1000,
  \dodoi{10.1046/j.1365-8711.2003.06897.x}

\bibitem[{{Burke} {et~al.}(2012){Burke}, {Collins}, {Stott}, \&
  {Hilton}}]{burke2012}
{Burke}, C., {Collins}, C.~A., {Stott}, J.~P., \& {Hilton}, M. 2012, \mnras,
  425, 2058, \dodoi{10.1111/j.1365-2966.2012.21555.x}

\bibitem[{{Burke} {et~al.}(2015){Burke}, {Hilton}, \& {Collins}}]{burke2015}
{Burke}, C., {Hilton}, M., \& {Collins}, C. 2015, \mnras, 449, 2353,
  \dodoi{10.1093/mnras/stv450}

\bibitem[{{Chabrier}(2003)}]{chabrier2003}
{Chabrier}, G. 2003, \pasp, 115, 763, \dodoi{10.1086/376392}

\bibitem[{{Chandar} {et~al.}(2021){Chandar}, {Mok}, {French}, {Smercina}, \&
  {Smith}}]{chandar2021}
{Chandar}, R., {Mok}, A., {French}, K.~D., {Smercina}, A., \& {Smith}, J.-D.~T.
  2021, \apj, 920, 105, \dodoi{10.3847/1538-4357/ac0c19}

\bibitem[{{Chen} {et~al.}(2022){Chen}, {Zu}, {Shao}, \& {Shan}}]{chen2022}
{Chen}, X., {Zu}, Y., {Shao}, Z., \& {Shan}, H. 2022, \mnras, 514, 2692,
  \dodoi{10.1093/mnras/stac1456}

\bibitem[{{Coccato} {et~al.}(2011){Coccato}, {Gerhard}, {Arnaboldi}, \&
  {Ventimiglia}}]{coccato2011}
{Coccato}, L., {Gerhard}, O., {Arnaboldi}, M., \& {Ventimiglia}, G. 2011, \aap,
  533, A138, \dodoi{10.1051/0004-6361/201117546}

\bibitem[{{Coe} {et~al.}(2019){Coe}, {Salmon}, {Brada{\v{c}}}, {Bradley},
  {Sharon}, {Zitrin}, {Acebron}, {Cerny}, {Cibirka}, {Strait},
  {Paterno-Mahler}, {Mahler}, {Avila}, {Ogaz}, {Huang}, {Pelliccia}, {Stark},
  {Mainali}, {Oesch}, {Trenti}, {Carrasco}, {Dawson}, {Rodney}, {Strolger},
  {Riess}, {Jones}, {Frye}, {Czakon}, {Umetsu}, {Vulcani}, {Graur}, {Jha},
  {Graham}, {Molino}, {Nonino}, {Hjorth}, {Selsing}, {Christensen},
  {Kikuchihara}, {Ouchi}, {Oguri}, {Welch}, {Lemaux}, {Andrade-Santos}, {Hoag},
  {Johnson}, {Peterson}, {Past}, {Fox}, {Agulli}, {Livermore}, {Ryan}, {Lam},
  {Sendra-Server}, {Toft}, {Lovisari}, \& {Su}}]{coe2019}
{Coe}, D., {Salmon}, B., {Brada{\v{c}}}, M., {et~al.} 2019, \apj, 884, 85,
  \dodoi{10.3847/1538-4357/ab412b}

\bibitem[{{Contini}(2021)}]{contini_review2021}
{Contini}, E. 2021, Galaxies, 9, 60, \dodoi{10.3390/galaxies9030060}

\bibitem[{{Contini} {et~al.}(2014){Contini}, {De Lucia}, {Villalobos}, \&
  {Borgani}}]{contini2014}
{Contini}, E., {De Lucia}, G., {Villalobos}, {\'A}., \& {Borgani}, S. 2014,
  \mnras, 437, 3787, \dodoi{10.1093/mnras/stt2174}

\bibitem[{{Contini} {et~al.}(2023{\natexlab{a}}){Contini}, {Jeon}, {Rhee},
  {Han}, \& {Yi}}]{contini2023}
{Contini}, E., {Jeon}, S., {Rhee}, J., {Han}, S., \& {Yi}, S.~K.
  2023{\natexlab{a}}, arXiv e-prints, arXiv:2310.03263,
  \dodoi{10.48550/arXiv.2310.03263}

\bibitem[{{Contini} {et~al.}(2023{\natexlab{b}}){Contini}, {Rhee}, {Han},
  {Jeon}, \& {Yi}}]{contini2023b}
{Contini}, E., {Rhee}, J., {Han}, S., {Jeon}, S., \& {Yi}, S.~K.
  2023{\natexlab{b}}, arXiv e-prints, arXiv:2310.20135,
  \dodoi{10.48550/arXiv.2310.20135}

\bibitem[{{de Oliveira} {et~al.}(2022){de Oliveira}, {Jim{\'e}nez-Teja}, \&
  {Dupke}}]{deoliveira2022}
{de Oliveira}, N. O.~L., {Jim{\'e}nez-Teja}, Y., \& {Dupke}, R. 2022, \mnras,
  512, 1916, \dodoi{10.1093/mnras/stac407}

\bibitem[{{Della Ceca} {et~al.}(2000){Della Ceca}, {Scaramella}, {Gioia},
  {Rosati}, {Fiore}, \& {Squires}}]{dellaceca2000}
{Della Ceca}, R., {Scaramella}, R., {Gioia}, I.~M., {et~al.} 2000, \aap, 353,
  498, \dodoi{10.48550/arXiv.astro-ph/9910489}

\bibitem[{{DeMaio} {et~al.}(2015){DeMaio}, {Gonzalez}, {Zabludoff}, {Zaritsky},
  \& {Brada{\v{c}}}}]{demaio2015}
{DeMaio}, T., {Gonzalez}, A.~H., {Zabludoff}, A., {Zaritsky}, D., \&
  {Brada{\v{c}}}, M. 2015, \mnras, 448, 1162, \dodoi{10.1093/mnras/stv033}

\bibitem[{{DeMaio} {et~al.}(2018){DeMaio}, {Gonzalez}, {Zabludoff}, {Zaritsky},
  {Connor}, {Donahue}, \& {Mulchaey}}]{demaio2018}
{DeMaio}, T., {Gonzalez}, A.~H., {Zabludoff}, A., {et~al.} 2018, \mnras, 474,
  3009, \dodoi{10.1093/mnras/stx2946}

\bibitem[{{Demarco} {et~al.}(2005){Demarco}, {Rosati}, {Lidman}, {Homeier},
  {Scannapieco}, {Ben{\'\i}tez}, {Mainieri}, {Nonino}, {Girardi}, {Stanford},
  {Tozzi}, {Borgani}, {Silk}, {Squires}, \& {Broadhurst}}]{demarco2005}
{Demarco}, R., {Rosati}, P., {Lidman}, C., {et~al.} 2005, \aap, 432, 381,
  \dodoi{10.1051/0004-6361:20041931}

\bibitem[{{den Brok} {et~al.}(2011){den Brok}, {Peletier}, {Valentijn},
  {Balcells}, {Carter}, {Erwin}, {Ferguson}, {Goudfrooij}, {Graham}, {Hammer},
  {Lucey}, {Trentham}, {Guzm{\'a}n}, {Hoyos}, {Verdoes Kleijn}, {Jogee},
  {Karick}, {Marinova}, {Mouhcine}, \& {Weinzirl}}]{denbrok2011}
{den Brok}, M., {Peletier}, R.~F., {Valentijn}, E.~A., {et~al.} 2011, \mnras,
  414, 3052, \dodoi{10.1111/j.1365-2966.2011.18606.x}

\bibitem[{{Diego} {et~al.}(2023){Diego}, {Meena}, {Adams}, {Broadhurst}, {Dai},
  {Coe}, {Frye}, {Kelly}, {Koekemoer}, {Pascale}, {Willner}, {Zackrisson},
  {Zitrin}, {Windhorst}, {Cohen}, {Jansen}, {Summers}, {Tompkins}, {Conselice},
  {Driver}, {Yan}, {Grogin}, {Marshall}, {Pirzkal}, {Robotham}, {Ryan},
  {Willmer}, {Bradley}, {Caminha}, {Caputi}, {Carleton}, \&
  {Kamieneski}}]{diego2023}
{Diego}, J.~M., {Meena}, A.~K., {Adams}, N.~J., {et~al.} 2023, \aap, 672, A3,
  \dodoi{10.1051/0004-6361/202245238}

\bibitem[{{Dupke} {et~al.}(2022){Dupke}, {Jimenez-teja}, {Su}, {Carrasco},
  {Koekemoer}, {Batalha}, {Johnson}, {Irwin}, {Miller}, {Dimauro}, {De
  Oliveira}, \& {Vilchez}}]{dupke2022}
{Dupke}, R.~A., {Jimenez-teja}, Y., {Su}, Y., {et~al.} 2022, arXiv e-prints,
  arXiv:2207.00603.
\newblock \doarXiv{2207.00603}

\bibitem[{{Ebeling} {et~al.}(2000){Ebeling}, {Jones}, {Perlman}, {Scharf},
  {Horner}, {Wegner}, {Malkan}, {Fairley}, \& {Mullis}}]{ebeling2000}
{Ebeling}, H., {Jones}, L.~R., {Perlman}, E., {et~al.} 2000, \apj, 534, 133,
  \dodoi{10.1086/308729}

\bibitem[{{Edwards} {et~al.}(2016){Edwards}, {Alpert}, {Trierweiler},
  {Abraham}, \& {Beizer}}]{edwards2016}
{Edwards}, L.~O.~V., {Alpert}, H.~S., {Trierweiler}, I.~L., {Abraham}, T., \&
  {Beizer}, V.~G. 2016, \mnras, 461, 230, \dodoi{10.1093/mnras/stw1314}

\bibitem[{{Elbaz} {et~al.}(2007){Elbaz}, {Daddi}, {Le Borgne}, {Dickinson},
  {Alexander}, {Chary}, {Starck}, {Brandt}, {Kitzbichler}, {MacDonald},
  {Nonino}, {Popesso}, {Stern}, \& {Vanzella}}]{elbaz2007}
{Elbaz}, D., {Daddi}, E., {Le Borgne}, D., {et~al.} 2007, \aap, 468, 33,
  \dodoi{10.1051/0004-6361:20077525}

\bibitem[{{Feldmeier} {et~al.}(2002){Feldmeier}, {Mihos}, {Morrison}, {Rodney},
  \& {Harding}}]{feldmeier2002}
{Feldmeier}, J.~J., {Mihos}, J.~C., {Morrison}, H.~L., {Rodney}, S.~A., \&
  {Harding}, P. 2002, \apj, 575, 779, \dodoi{10.1086/341472}

\bibitem[{{Furnell} {et~al.}(2021){Furnell}, {Collins}, {Kelvin}, {Baldry},
  {James}, {Manolopoulou}, {Mann}, {Giles}, {Bermeo}, {Hilton}, {Wilkinson},
  {Romer}, {Vergara}, {Bhargava}, {Stott}, {Mayers}, \& {Viana}}]{furnell2021}
{Furnell}, K.~E., {Collins}, C.~A., {Kelvin}, L.~S., {et~al.} 2021, \mnras,
  502, 2419, \dodoi{10.1093/mnras/stab065}

\bibitem[{{Gerhard} {et~al.}(2005){Gerhard}, {Arnaboldi}, {Freeman},
  {Kashikawa}, {Okamura}, \& {Yasuda}}]{gerhard2005}
{Gerhard}, O., {Arnaboldi}, M., {Freeman}, K.~C., {et~al.} 2005, \apjl, 621,
  L93, \dodoi{10.1086/429221}

\bibitem[{{Girardi} {et~al.}(2005){Girardi}, {Demarco}, {Rosati}, \&
  {Borgani}}]{girardi2005}
{Girardi}, M., {Demarco}, R., {Rosati}, P., \& {Borgani}, S. 2005, \aap, 442,
  29, \dodoi{10.1051/0004-6361:20053232}

\bibitem[{{Golden-Marx} {et~al.}(2023){Golden-Marx}, {Zhang}, {Ogando},
  {Allam}, {Tucker}, {Miller}, {Hilton}, {Mutlu-Pakdil}, {Abbott}, {Aguena},
  {Alves}, {Andrade-Oliveira}, {Annis}, {Bacon}, {Bertin}, {Bocquet}, {Brooks},
  {Burke}, {Carnero Rosell}, {Carrasco Kind}, {Castander}, {Conselice},
  {Costanzi}, {da Costa}, {Pereira}, {De Vicente}, {Desai}, {Doel}, {Everett},
  {Ferrero}, {Flaugher}, {Frieman}, {Garc{\'\i}a-Bellido}, {Gerdes}, {Gruen},
  {Gruendl}, {Gutierrez}, {Hinton}, {Hollowood}, {Honscheid}, {James}, {Kuehn},
  {Kuropatkin}, {Lahav}, {Marshall}, {Melchior}, {Mena-Fern{\'a}ndez},
  {Miquel}, {Mohr}, {Palmese}, {Paz-Chinch{\'o}n}, {Pieres}, {Plazas
  Malag{\'o}n}, {Prat}, {Raveri}, {Rodriguez-Monroy}, {Romer}, {Sanchez},
  {Scarpine}, {Sevilla-Noarbe}, {Sif{\'o}n}, {Smith}, {Suchyta}, {Swanson},
  {Tarle}, {Vincenzi}, {Weaverdyck}, {Yanny}, \& {DES
  Collaboration}}]{golden-marx2023}
{Golden-Marx}, J.~B., {Zhang}, Y., {Ogando}, R.~L.~C., {et~al.} 2023, \mnras,
  521, 478, \dodoi{10.1093/mnras/stad469}

\bibitem[{{Gonzalez} {et~al.}(2021){Gonzalez}, {George}, {Connor}, {Deason},
  {Donahue}, {Montes}, {Zabludoff}, \& {Zaritsky}}]{gonzalez2021}
{Gonzalez}, A.~H., {George}, T., {Connor}, T., {et~al.} 2021, \mnras, 507, 963,
  \dodoi{10.1093/mnras/stab2117}

\bibitem[{{Gonzalez} {et~al.}(2005){Gonzalez}, {Zabludoff}, \&
  {Zaritsky}}]{gonzalez2005}
{Gonzalez}, A.~H., {Zabludoff}, A.~I., \& {Zaritsky}, D. 2005, \apj, 618, 195,
  \dodoi{10.1086/425896}

\bibitem[{{Gonz{\'a}lez Delgado} {et~al.}(2015){Gonz{\'a}lez Delgado},
  {Garc{\'\i}a-Benito}, {P{\'e}rez}, {Cid Fernandes}, {de Amorim},
  {Cortijo-Ferrero}, {Lacerda}, {L{\'o}pez Fern{\'a}ndez}, {Vale-Asari},
  {S{\'a}nchez}, {Moll{\'a}}, {Ruiz-Lara}, {S{\'a}nchez-Bl{\'a}zquez},
  {Walcher}, {Alves}, {Aguerri}, {Bekerait{\'e}}, {Bland-Hawthorn}, {Galbany},
  {Gallazzi}, {Husemann}, {Iglesias-P{\'a}ramo}, {Kalinova},
  {L{\'o}pez-S{\'a}nchez}, {Marino}, {M{\'a}rquez}, {Masegosa}, {Mast},
  {M{\'e}ndez-Abreu}, {Mendoza}, {del Olmo}, {P{\'e}rez}, {Quirrenbach}, \&
  {Zibetti}}]{gonzalez-delgado2015}
{Gonz{\'a}lez Delgado}, R.~M., {Garc{\'\i}a-Benito}, R., {P{\'e}rez}, E.,
  {et~al.} 2015, \aap, 581, A103, \dodoi{10.1051/0004-6361/201525938}

\bibitem[{{Guennou} {et~al.}(2012){Guennou}, {Adami}, {Da Rocha}, {Durret},
  {Ulmer}, {Allam}, {Basa}, {Benoist}, {Biviano}, {Clowe}, {Gavazzi},
  {Halliday}, {Ilbert}, {Johnston}, {Just}, {Kron}, {Kubo}, {Le Brun},
  {Marshall}, {Mazure}, {Murphy}, {Pereira}, {Raba{\c{c}}a}, {Rostagni},
  {Rudnick}, {Russeil}, {Schrabback}, {Slezak}, {Tucker}, \&
  {Zaritsky}}]{guennou2012}
{Guennou}, L., {Adami}, C., {Da Rocha}, C., {et~al.} 2012, \aap, 537, A64,
  \dodoi{10.1051/0004-6361/201117482}

\bibitem[{{Gullieuszik} {et~al.}(2020){Gullieuszik}, {Poggianti}, {McGee},
  {Moretti}, {Vulcani}, {Tonnesen}, {Roediger}, {Jaff{\'e}}, {Fritz},
  {Franchetto}, {Omizzolo}, {Bettoni}, {Radovich}, \&
  {Wolter}}]{gullieuszik2020}
{Gullieuszik}, M., {Poggianti}, B.~M., {McGee}, S.~L., {et~al.} 2020, \apj,
  899, 13, \dodoi{10.3847/1538-4357/aba3cb}

\bibitem[{{Haigh} {et~al.}(2021){Haigh}, {Chamba}, {Venhola}, {Peletier},
  {Doorenbos}, {Watkins}, \& {Wilkinson}}]{haigh2021}
{Haigh}, C., {Chamba}, N., {Venhola}, A., {et~al.} 2021, \aap, 645, A107,
  \dodoi{10.1051/0004-6361/201936561}

\bibitem[{{Hilton} {et~al.}(2021){Hilton}, {Sif{\'o}n}, {Naess},
  {Madhavacheril}, {Oguri}, {Rozo}, {Rykoff}, {Abbott}, {Adhikari}, {Aguena},
  {Aiola}, {Allam}, {Amodeo}, {Amon}, {Annis}, {Ansarinejad}, {Aros-Bunster},
  {Austermann}, {Avila}, {Bacon}, {Battaglia}, {Beall}, {Becker}, {Bernstein},
  {Bertin}, {Bhandarkar}, {Bhargava}, {Bond}, {Brooks}, {Burke}, {Calabrese},
  {Carrasco Kind}, {Carretero}, {Choi}, {Choi}, {Conselice}, {da Costa},
  {Costanzi}, {Crichton}, {Crowley}, {D{\"u}nner}, {Denison}, {Devlin},
  {Dicker}, {Diehl}, {Dietrich}, {Doel}, {Duff}, {Duivenvoorden}, {Dunkley},
  {Everett}, {Ferraro}, {Ferrero}, {Fert{\'e}}, {Flaugher}, {Frieman},
  {Gallardo}, {Garc{\'\i}a-Bellido}, {Gaztanaga}, {Gerdes}, {Giles}, {Golec},
  {Gralla}, {Grandis}, {Gruen}, {Gruendl}, {Gschwend}, {Gutierrez}, {Han},
  {Hartley}, {Hasselfield}, {Hill}, {Hilton}, {Hincks}, {Hinton}, {Ho},
  {Honscheid}, {Hoyle}, {Hubmayr}, {Huffenberger}, {Hughes}, {Jaelani}, {Jain},
  {James}, {Jeltema}, {Kent}, {Knowles}, {Koopman}, {Kuehn}, {Lahav}, {Lima},
  {Lin}, {Lokken}, {Loubser}, {MacCrann}, {Maia}, {Marriage}, {Martin},
  {McMahon}, {Melchior}, {Menanteau}, {Miquel}, {Miyatake}, {Moodley},
  {Morgan}, {Mroczkowski}, {Nati}, {Newburgh}, {Niemack}, {Nishizawa},
  {Ogando}, {Orlowski-Scherer}, {Page}, {Palmese}, {Partridge},
  {Paz-Chinch{\'o}n}, {Phakathi}, {Plazas}, {Robertson}, {Romer}, {Carnero
  Rosell}, {Salatino}, {Sanchez}, {Schaan}, {Schillaci}, {Sehgal}, {Serrano},
  {Shin}, {Simon}, {Smith}, {Soares-Santos}, {Spergel}, {Staggs}, {Storer},
  {Suchyta}, {Swanson}, {Tarle}, {Thomas}, {To}, {Trac}, {Ullom}, {Vale}, {Van
  Lanen}, {Vavagiakis}, {De Vicente}, {Wilkinson}, {Wollack}, {Xu}, \&
  {Zhang}}]{hilton2021}
{Hilton}, M., {Sif{\'o}n}, C., {Naess}, S., {et~al.} 2021, \apjs, 253, 3,
  \dodoi{10.3847/1538-4365/abd023}

\bibitem[{{Hlavacek-Larrondo} {et~al.}(2020){Hlavacek-Larrondo}, {Rhea},
  {Webb}, {McDonald}, {Muzzin}, {Wilson}, {Finner}, {Valin}, {Bonaventura},
  {Cooper}, {Fabian}, {Gendron-Marsolais}, {Jee}, {Lidman}, {Mezcua}, {Noble},
  {Russell}, {Surace}, {Trudeau}, \& {Yee}}]{hlavacek-larrondo2020}
{Hlavacek-Larrondo}, J., {Rhea}, C.~L., {Webb}, T., {et~al.} 2020, \apjl, 898,
  L50, \dodoi{10.3847/2041-8213/ab9ca5}

\bibitem[{{Huo} {et~al.}(2004){Huo}, {Xue}, {Xu}, {Squires}, \&
  {Rosati}}]{huo2004}
{Huo}, Z.-Y., {Xue}, S.-J., {Xu}, H., {Squires}, G., \& {Rosati}, P. 2004, \aj,
  127, 1263, \dodoi{10.1086/381949}

\bibitem[{{Jee} {et~al.}(2005){Jee}, {White}, {Ben{\'\i}tez}, {Ford},
  {Blakeslee}, {Rosati}, {Demarco}, \& {Illingworth}}]{jee2005}
{Jee}, M.~J., {White}, R.~L., {Ben{\'\i}tez}, N., {et~al.} 2005, \apj, 618, 46,
  \dodoi{10.1086/425912}

\bibitem[{{Jim{\'e}nez-Teja} \& {Ben{\'\i}tez}(2012)}]{jimenez-teja2012}
{Jim{\'e}nez-Teja}, Y., \& {Ben{\'\i}tez}, N. 2012, \apj, 745, 150,
  \dodoi{10.1088/0004-637X/745/2/150}

\bibitem[{{Jim{\'e}nez-Teja} \& {Dupke}(2016)}]{jimenez-teja2016}
{Jim{\'e}nez-Teja}, Y., \& {Dupke}, R. 2016, \apj, 820, 49,
  \dodoi{10.3847/0004-637X/820/1/49}

\bibitem[{{Jim{\'e}nez-Teja} {et~al.}(2023){Jim{\'e}nez-Teja}, {Dupke},
  {Lopes}, \& {V{\'\i}lchez}}]{jimenez-teja2023}
{Jim{\'e}nez-Teja}, Y., {Dupke}, R.~A., {Lopes}, P.~A.~A., \& {V{\'\i}lchez},
  J.~M. 2023, \aap, 676, A39, \dodoi{10.1051/0004-6361/202346580}

\bibitem[{{Jim{\'e}nez-Teja} {et~al.}(2021){Jim{\'e}nez-Teja}, {V{\'\i}lchez},
  {Dupke}, {Lopes}, {de Oliveira}, \& {Coe}}]{jimenez-teja2021}
{Jim{\'e}nez-Teja}, Y., {V{\'\i}lchez}, J.~M., {Dupke}, R.~A., {et~al.} 2021,
  arXiv e-prints, arXiv:2109.04485.
\newblock \doarXiv{2109.04485}

\bibitem[{{Jim{\'e}nez-Teja} {et~al.}(2018){Jim{\'e}nez-Teja}, {Dupke},
  {Ben{\'\i}tez}, {Koekemoer}, {Zitrin}, {Umetsu}, {Ziegler}, {Frye}, {Ford},
  {Bouwens}, {Bradley}, {Broadhurst}, {Coe}, {Donahue}, {Graves}, {Grillo},
  {Infante}, {Jouvel}, {Kelson}, {Lahav}, {Lazkoz}, {Lemze}, {Maoz},
  {Medezinski}, {Melchior}, {Meneghetti}, {Mercurio}, {Merten}, {Molino},
  {Moustakas}, {Nonino}, {Ogaz}, {Riess}, {Rosati}, {Sayers}, {Seitz}, \&
  {Zheng}}]{jimenez-teja2018}
{Jim{\'e}nez-Teja}, Y., {Dupke}, R., {Ben{\'\i}tez}, N., {et~al.} 2018, \apj,
  857, 79, \dodoi{10.3847/1538-4357/aab70f}

\bibitem[{{Jim{\'e}nez-Teja} {et~al.}(2019){Jim{\'e}nez-Teja}, {Dupke}, {Lopes
  de Oliveira}, {Xavier}, {Coelho}, {Chies-Santos}, {L{\'o}pez-Sanjuan},
  {Alvarez-Candal}, {Costa-Duarte}, {Telles}, {Hernandez-Jimenez},
  {Ben{\'\i}tez}, {Alcaniz}, {Cenarro}, {Crist{\'o}bal-Hornillos},
  {Ederoclite}, {Mar{\'\i}n-Franch}, {Mendes de Oliveira}, {Moles},
  {Sodr{\'e}}, {Varela}, \& {V{\'a}zquez Rami{\'o}}}]{jimenez-teja2019}
{Jim{\'e}nez-Teja}, Y., {Dupke}, R.~A., {Lopes de Oliveira}, R., {et~al.} 2019,
  \aap, 622, A183, \dodoi{10.1051/0004-6361/201833547}

\bibitem[{{Joo} \& {Jee}(2023)}]{joo2023}
{Joo}, H., \& {Jee}, M.~J. 2023, \nat, 613, 37,
  \dodoi{10.1038/s41586-022-05396-4}

\bibitem[{{J{\o}rgensen} {et~al.}(2005){J{\o}rgensen}, {Bergmann}, {Davies},
  {Barr}, {Takamiya}, \& {Crampton}}]{jorgensen2005}
{J{\o}rgensen}, I., {Bergmann}, M., {Davies}, R., {et~al.} 2005, \aj, 129,
  1249, \dodoi{10.1086/427857}

\bibitem[{{Joy} {et~al.}(2001){Joy}, {LaRoque}, {Grego}, {Carlstrom}, {Dawson},
  {Ebeling}, {Holzapfel}, {Nagai}, \& {Reese}}]{joy2001}
{Joy}, M., {LaRoque}, S., {Grego}, L., {et~al.} 2001, \apjl, 551, L1,
  \dodoi{10.1086/319833}

\bibitem[{{Kelvin} {et~al.}(2023){Kelvin}, {Hasan}, \& {Tyson}}]{kelvin2023}
{Kelvin}, L.~S., {Hasan}, I., \& {Tyson}, J.~A. 2023, \mnras, 520, 2484,
  \dodoi{10.1093/mnras/stad180}

\bibitem[{{Kim} {et~al.}(2021){Kim}, {Jee}, {Hughes}, {Yoon}, {HyeongHan},
  {Menanteau}, {Sif{\'o}n}, {Hovey}, \& {Arunachalam}}]{kim2021}
{Kim}, J., {Jee}, M.~J., {Hughes}, J.~P., {et~al.} 2021, \apj, 923, 101,
  \dodoi{10.3847/1538-4357/ac294f}

\bibitem[{{Kluge} {et~al.}(2021){Kluge}, {Bender}, {Riffeser}, {Goessl},
  {Hopp}, {Schmidt}, \& {Ries}}]{kluge2021}
{Kluge}, M., {Bender}, R., {Riffeser}, A., {et~al.} 2021, \apjs, 252, 27,
  \dodoi{10.3847/1538-4365/abcda6}

\bibitem[{{Kluge} {et~al.}(2020){Kluge}, {Neureiter}, {Riffeser}, {Bender},
  {Goessl}, {Hopp}, {Schmidt}, {Ries}, \& {Brosch}}]{kluge2020}
{Kluge}, M., {Neureiter}, B., {Riffeser}, A., {et~al.} 2020, \apjs, 247, 43,
  \dodoi{10.3847/1538-4365/ab733b}

\bibitem[{{Koekemoer} \& {et al.}(2002)}]{koekemoer2002}
{Koekemoer}, A.~M., \& {et al.} 2002, {HST Dither Handbook}

\bibitem[{{Krick} \& {Bernstein}(2007)}]{krick2007}
{Krick}, J.~E., \& {Bernstein}, R.~A. 2007, \aj, 134, 466,
  \dodoi{10.1086/518787}

\bibitem[{{Krick} {et~al.}(2006){Krick}, {Bernstein}, \&
  {Pimbblet}}]{krick2006}
{Krick}, J.~E., {Bernstein}, R.~A., \& {Pimbblet}, K.~A. 2006, \aj, 131, 168,
  \dodoi{10.1086/498269}

\bibitem[{{Lee} {et~al.}(2022){Lee}, {Bae}, \& {Jang}}]{lee2022}
{Lee}, M.~G., {Bae}, J.~H., \& {Jang}, I.~S. 2022, \apjl, 940, L19,
  \dodoi{10.3847/2041-8213/ac990b}

\bibitem[{{Lin} \& {Mohr}(2004)}]{lin2004}
{Lin}, Y.-T., \& {Mohr}, J.~J. 2004, \apj, 617, 879, \dodoi{10.1086/425412}

\bibitem[{{Lindner} {et~al.}(2014){Lindner}, {Baker}, {Hughes}, {Battaglia},
  {Gupta}, {Knowles}, {Marriage}, {Menanteau}, {Moodley}, {Reese}, \&
  {Srianand}}]{lindner2014}
{Lindner}, R.~R., {Baker}, A.~J., {Hughes}, J.~P., {et~al.} 2014, \apj, 786,
  49, \dodoi{10.1088/0004-637X/786/1/49}

\bibitem[{{Lopes} \& {Ribeiro}(2020)}]{lop20}
{Lopes}, P. A.~A., \& {Ribeiro}, A. L.~B. 2020, \mnras, 493, 3429,
  \dodoi{10.1093/mnras/staa486}

\bibitem[{{Mancone} \& {Gonzalez}(2012)}]{mancone2012}
{Mancone}, C., \& {Gonzalez}, A. 2012, {EzGal: A Flexible Interface for Stellar
  Population Synthesis Models}, Astrophysics Source Code Library, record
  ascl:1208.021.
\newblock \doeprint{1208.021}

\bibitem[{{Marriage} {et~al.}(2011){Marriage}, {Acquaviva}, {Ade}, {Aguirre},
  {Amiri}, {Appel}, {Barrientos}, {Battistelli}, {Bond}, {Brown}, {Burger},
  {Chervenak}, {Das}, {Devlin}, {Dicker}, {Bertrand Doriese}, {Dunkley},
  {D{\"u}nner}, {Essinger-Hileman}, {Fisher}, {Fowler}, {Hajian}, {Halpern},
  {Hasselfield}, {Hern{\'a}ndez-Monteagudo}, {Hilton}, {Hilton}, {Hincks},
  {Hlozek}, {Huffenberger}, {Handel Hughes}, {Hughes}, {Infante}, {Irwin},
  {Baptiste Juin}, {Kaul}, {Klein}, {Kosowsky}, {Lau}, {Limon}, {Lin},
  {Lupton}, {Marsden}, {Martocci}, {Mauskopf}, {Menanteau}, {Moodley},
  {Moseley}, {Netterfield}, {Niemack}, {Nolta}, {Page}, {Parker}, {Partridge},
  {Quintana}, {Reese}, {Reid}, {Sehgal}, {Sherwin}, {Sievers}, {Spergel},
  {Staggs}, {Swetz}, {Switzer}, {Thornton}, {Trac}, {Tucker}, {Warne},
  {Wilson}, {Wollack}, \& {Zhao}}]{marriage2011}
{Marriage}, T.~A., {Acquaviva}, V., {Ade}, P. A.~R., {et~al.} 2011, \apj, 737,
  61, \dodoi{10.1088/0004-637X/737/2/61}

\bibitem[{{Maughan} {et~al.}(2003){Maughan}, {Jones}, {Ebeling}, {Perlman},
  {Rosati}, {Frye}, \& {Mullis}}]{maughan2003}
{Maughan}, B.~J., {Jones}, L.~R., {Ebeling}, H., {et~al.} 2003, \apj, 587, 589,
  \dodoi{10.1086/368151}

\bibitem[{{Melnick} {et~al.}(2012){Melnick}, {Giraud}, {Toledo}, {Selman}, \&
  {Quintana}}]{melnick2012}
{Melnick}, J., {Giraud}, E., {Toledo}, I., {Selman}, F., \& {Quintana}, H.
  2012, \mnras, 427, 850, \dodoi{10.1111/j.1365-2966.2012.21924.x}

\bibitem[{{Menanteau} {et~al.}(2010){Menanteau}, {Gonz{\'a}lez}, {Juin},
  {Marriage}, {Reese}, {Acquaviva}, {Aguirre}, {Appel}, {Baker}, {Barrientos},
  {Battistelli}, {Bond}, {Das}, {Deshpande}, {Devlin}, {Dicker}, {Dunkley},
  {D{\"u}nner}, {Essinger-Hileman}, {Fowler}, {Hajian}, {Halpern},
  {Hasselfield}, {Hern{\'a}ndez-Monteagudo}, {Hilton}, {Hincks}, {Hlozek},
  {Huffenberger}, {Hughes}, {Infante}, {Irwin}, {Klein}, {Kosowsky}, {Lin},
  {Marsden}, {Moodley}, {Niemack}, {Nolta}, {Page}, {Parker}, {Partridge},
  {Sehgal}, {Sievers}, {Spergel}, {Staggs}, {Swetz}, {Switzer}, {Thornton},
  {Trac}, {Warne}, \& {Wollack}}]{menanteau2010}
{Menanteau}, F., {Gonz{\'a}lez}, J., {Juin}, J.-B., {et~al.} 2010, \apj, 723,
  1523, \dodoi{10.1088/0004-637X/723/2/1523}

\bibitem[{{Menanteau} {et~al.}(2012){Menanteau}, {Hughes}, {Sif{\'o}n},
  {Hilton}, {Gonz{\'a}lez}, {Infante}, {Barrientos}, {Baker}, {Bond}, {Das},
  {Devlin}, {Dunkley}, {Hajian}, {Hincks}, {Kosowsky}, {Marsden}, {Marriage},
  {Moodley}, {Niemack}, {Nolta}, {Page}, {Reese}, {Sehgal}, {Sievers},
  {Spergel}, {Staggs}, \& {Wollack}}]{menanteau2012}
{Menanteau}, F., {Hughes}, J.~P., {Sif{\'o}n}, C., {et~al.} 2012, \apj, 748, 7,
  \dodoi{10.1088/0004-637X/748/1/7}

\bibitem[{{Mihos} {et~al.}(2005){Mihos}, {Harding}, {Feldmeier}, \&
  {Morrison}}]{mihos2005}
{Mihos}, J.~C., {Harding}, P., {Feldmeier}, J., \& {Morrison}, H. 2005, \apjl,
  631, L41, \dodoi{10.1086/497030}

\bibitem[{{Mihos} {et~al.}(2017){Mihos}, {Harding}, {Feldmeier}, {Rudick},
  {Janowiecki}, {Morrison}, {Slater}, \& {Watkins}}]{mihos2017}
{Mihos}, J.~C., {Harding}, P., {Feldmeier}, J.~J., {et~al.} 2017, \apj, 834,
  16, \dodoi{10.3847/1538-4357/834/1/16}

\bibitem[{{Molnar} \& {Broadhurst}(2015)}]{molnar2015}
{Molnar}, S.~M., \& {Broadhurst}, T. 2015, \apj, 800, 37,
  \dodoi{10.1088/0004-637X/800/1/37}

\bibitem[{{Molnar} {et~al.}(2012){Molnar}, {Hearn}, \& {Stadel}}]{molnar2012}
{Molnar}, S.~M., {Hearn}, N.~C., \& {Stadel}, J.~G. 2012, \apj, 748, 45,
  \dodoi{10.1088/0004-637X/748/1/45}

\bibitem[{{Montes}(2022)}]{montes_review2022}
{Montes}, M. 2022, Nature Astronomy, 6, 308, \dodoi{10.1038/s41550-022-01616-z}

\bibitem[{{Montes} \& {Trujillo}(2014)}]{montes2014}
{Montes}, M., \& {Trujillo}, I. 2014, \apj, 794, 137,
  \dodoi{10.1088/0004-637X/794/2/137}

\bibitem[{{Montes} \& {Trujillo}(2018)}]{montes2018}
---. 2018, \mnras, 474, 917, \dodoi{10.1093/mnras/stx2847}

\bibitem[{{Montes} \& {Trujillo}(2022)}]{montes2022}
---. 2022, \apjl, 940, L51, \dodoi{10.3847/2041-8213/ac98c5}

\bibitem[{{Morishita} {et~al.}(2017){Morishita}, {Abramson}, {Treu}, {Schmidt},
  {Vulcani}, \& {Wang}}]{morishita2017}
{Morishita}, T., {Abramson}, L.~E., {Treu}, T., {et~al.} 2017, \apj, 846, 139,
  \dodoi{10.3847/1538-4357/aa8403}

\bibitem[{{Pierini} {et~al.}(2008){Pierini}, {Zibetti}, {Braglia},
  {B{\"o}hringer}, {Finoguenov}, {Lynam}, \& {Zhang}}]{pierini2008}
{Pierini}, D., {Zibetti}, S., {Braglia}, F., {et~al.} 2008, \aap, 483, 727,
  \dodoi{10.1051/0004-6361:200809400}

\bibitem[{{Postman} {et~al.}(2012){Postman}, {Coe}, {Ben{\'\i}tez}, {Bradley},
  {Broadhurst}, {Donahue}, {Ford}, {Graur}, {Graves}, {Jouvel}, {Koekemoer},
  {Lemze}, {Medezinski}, {Molino}, {Moustakas}, {Ogaz}, {Riess}, {Rodney},
  {Rosati}, {Umetsu}, {Zheng}, {Zitrin}, {Bartelmann}, {Bouwens}, {Czakon},
  {Golwala}, {Host}, {Infante}, {Jha}, {Jimenez-Teja}, {Kelson}, {Lahav},
  {Lazkoz}, {Maoz}, {McCully}, {Melchior}, {Meneghetti}, {Merten}, {Moustakas},
  {Nonino}, {Patel}, {Reg{\"o}s}, {Sayers}, {Seitz}, \& {Van der
  Wel}}]{postman2012}
{Postman}, M., {Coe}, D., {Ben{\'\i}tez}, N., {et~al.} 2012, \apjs, 199, 25,
  \dodoi{10.1088/0067-0049/199/2/25}

\bibitem[{{Ragusa} {et~al.}(2023){Ragusa}, {Iodice}, {Spavone}, {Montes},
  {Forbes}, {Brough}, {Mirabile}, {Cantiello}, {Paolillo}, \&
  {Schipani}}]{ragusa2023}
{Ragusa}, R., {Iodice}, E., {Spavone}, M., {et~al.} 2023, \aap, 670, L20,
  \dodoi{10.1051/0004-6361/202245530}

\bibitem[{{Raj} {et~al.}(2020){Raj}, {Iodice}, {Napolitano}, {Hilker},
  {Spavone}, {Peletier}, {Su}, {Falc{\'o}n-Barroso}, {van de Ven}, {Cantiello},
  {Kleiner}, {Venhola}, {Mieske}, {Paolillo}, {Capaccioli}, \&
  {Schipani}}]{raj2020}
{Raj}, M.~A., {Iodice}, E., {Napolitano}, N.~R., {et~al.} 2020, \aap, 640,
  A137, \dodoi{10.1051/0004-6361/202038043}

\bibitem[{{Rom{\'a}n} {et~al.}(2020){Rom{\'a}n}, {Trujillo}, \&
  {Montes}}]{roman2020}
{Rom{\'a}n}, J., {Trujillo}, I., \& {Montes}, M. 2020, \aap, 644, A42,
  \dodoi{10.1051/0004-6361/201936111}

\bibitem[{{Romer} {et~al.}(2000){Romer}, {Nichol}, {Holden}, {Ulmer}, {Pildis},
  {Merrelli}, {Adami}, {Burke}, {Collins}, {Metevier}, {Kron}, \&
  {Commons}}]{romer2000}
{Romer}, A.~K., {Nichol}, R.~C., {Holden}, B.~P., {et~al.} 2000, \apjs, 126,
  209, \dodoi{10.1086/313302}

\bibitem[{{Rosati} {et~al.}(1998){Rosati}, {Della Ceca}, {Norman}, \&
  {Giacconi}}]{rosati1998}
{Rosati}, P., {Della Ceca}, R., {Norman}, C., \& {Giacconi}, R. 1998, \apjl,
  492, L21, \dodoi{10.1086/311085}

\bibitem[{{Rudick} {et~al.}(2011){Rudick}, {Mihos}, \& {McBride}}]{rudick2011}
{Rudick}, C.~S., {Mihos}, J.~C., \& {McBride}, C.~K. 2011, \apj, 732, 48,
  \dodoi{10.1088/0004-637X/732/1/48}

\bibitem[{{Sayers} {et~al.}(2013){Sayers}, {Czakon}, {Mantz}, {Golwala},
  {Ameglio}, {Downes}, {Koch}, {Lin}, {Maughan}, {Molnar}, {Moustakas},
  {Mroczkowski}, {Pierpaoli}, {Shitanishi}, {Siegel}, {Umetsu}, \& {Van der
  Pyl}}]{sayers2013}
{Sayers}, J., {Czakon}, N.~G., {Mantz}, A., {et~al.} 2013, \apj, 768, 177,
  \dodoi{10.1088/0004-637X/768/2/177}

\bibitem[{{Tanaka} {et~al.}(2006){Tanaka}, {Kodama}, {Arimoto}, \&
  {Tanaka}}]{tanaka2006}
{Tanaka}, M., {Kodama}, T., {Arimoto}, N., \& {Tanaka}, I. 2006, \mnras, 365,
  1392, \dodoi{10.1111/j.1365-2966.2005.09841.x}

\bibitem[{{Toledo} {et~al.}(2011){Toledo}, {Melnick}, {Selman}, {Quintana},
  {Giraud}, \& {Zelaya}}]{toledo2011}
{Toledo}, I., {Melnick}, J., {Selman}, F., {et~al.} 2011, \mnras, 414, 602,
  \dodoi{10.1111/j.1365-2966.2011.18423.x}

\bibitem[{{Williams} {et~al.}(2007){Williams}, {Ciardullo}, {Durrell},
  {Vinciguerra}, {Feldmeier}, {Jacoby}, {Sigurdsson}, {von Hippel}, {Ferguson},
  {Tanvir}, {Arnaboldi}, {Gerhard}, {Aguerri}, \& {Freeman}}]{williams2007}
{Williams}, B.~F., {Ciardullo}, R., {Durrell}, P.~R., {et~al.} 2007, \apj, 656,
  756, \dodoi{10.1086/510149}

\bibitem[{{Yoo} {et~al.}(2021){Yoo}, {Ko}, {Kim}, \& {Kim}}]{yoo2021}
{Yoo}, J., {Ko}, J., {Kim}, J.-W., \& {Kim}, H. 2021, \mnras, 508, 2634,
  \dodoi{10.1093/mnras/stab2707}

\bibitem[{{Yuan} {et~al.}(2022){Yuan}, {Han}, \& {Wen}}]{yuan2022}
{Yuan}, Z.~S., {Han}, J.~L., \& {Wen}, Z.~L. 2022, \mnras, 513, 3013,
  \dodoi{10.1093/mnras/stac1037}

\bibitem[{{Zhang} {et~al.}(2019){Zhang}, {Yanny}, {Palmese}, {Gruen}, {To},
  {Rykoff}, {Leung}, {Collins}, {Hilton}, {Abbott}, {Annis}, {Avila}, {Bertin},
  {Brooks}, {Burke}, {Carnero Rosell}, {Carrasco Kind}, {Carretero}, {Cunha},
  {D'Andrea}, {da Costa}, {De Vicente}, {Desai}, {Diehl}, {Dietrich}, {Doel},
  {Drlica-Wagner}, {Eifler}, {Evrard}, {Flaugher}, {Fosalba}, {Frieman},
  {Garc{\'\i}a-Bellido}, {Gaztanaga}, {Gerdes}, {Gruendl}, {Gschwend},
  {Gutierrez}, {Hartley}, {Hollowood}, {Honscheid}, {Hoyle}, {James},
  {Jeltema}, {Kuehn}, {Kuropatkin}, {Li}, {Lima}, {Maia}, {March}, {Marshall},
  {Melchior}, {Menanteau}, {Miller}, {Miquel}, {Mohr}, {Ogando}, {Plazas},
  {Romer}, {Sanchez}, {Scarpine}, {Schubnell}, {Serrano}, {Sevilla-Noarbe},
  {Smith}, {Soares-Santos}, {Sobreira}, {Suchyta}, {Swanson}, {Tarle},
  {Thomas}, {Wester}, \& {DES Collaboration}}]{zhang2019}
{Zhang}, Y., {Yanny}, B., {Palmese}, A., {et~al.} 2019, \apj, 874, 165,
  \dodoi{10.3847/1538-4357/ab0dfd}

\bibitem[{{Zhang} {et~al.}(2023){Zhang}, {Golden-Marx}, {Ogando}, {Yanny},
  {Rykoff}, {Allam}, {Aguena}, {Bacon}, {Bocquet}, {Brooks}, {Carnero Rosell},
  {Carretero}, {Cheng}, {Conselice}, {Costanzi}, {da Costa}, {Pereira},
  {Davis}, {Desai}, {Diehl}, {Doel}, {Ferrero}, {Flaugher}, {Frieman}, {Gruen},
  {Gruendl}, {Hinton}, {Hollowood}, {Honscheid}, {James}, {Jeltema}, {Kuehn},
  {Kuropatkin}, {Lahav}, {Lee}, {Lima}, {Mena-Fern{\'a}ndez}, {Miquel},
  {Palmese}, {Pieres}, {Plazas Malag{\'o}n}, {Romer}, {Sanchez}, {Smith},
  {Suchyta}, {Tarle}, {To}, {Tucker}, \& {Weaverdyck}}]{zhang2023}
{Zhang}, Y., {Golden-Marx}, J.~B., {Ogando}, R. L.~C., {et~al.} 2023, arXiv
  e-prints, arXiv:2309.00671, \dodoi{10.48550/arXiv.2309.00671}

\bibitem[{{Zibetti} {et~al.}(2005){Zibetti}, {White}, {Schneider}, \&
  {Brinkmann}}]{zibetti2005}
{Zibetti}, S., {White}, S. D.~M., {Schneider}, D.~P., \& {Brinkmann}, J. 2005,
  \mnras, 358, 949, \dodoi{10.1111/j.1365-2966.2005.08817.x}

\end{thebibliography}
\bibliographystyle{aasjournal}

\end{document}